\documentclass[
twocolumn,
prb,preprintnumbers,amsmath,amssymb]{revtex4}

\usepackage{times}
\usepackage{physics}
\usepackage{graphicx}
\usepackage{amssymb}
\usepackage{amsthm}
\usepackage{amsmath}
\usepackage{dsfont}
\usepackage{bm}
\usepackage{mathrsfs}
\usepackage{bbold}
\usepackage{color}

\begin{document}

\title{Van der Waals chain: a simple model for Casimir forces in dielectrics}
\author{Helmut H\"{o}rner$^1$, Lukas M. Rachbauer$^1$, Stefan Rotter$^1$, and Ulf Leonhardt$^2$}
\affiliation{
\normalsize{
$^1$Institute for Theoretical Physics, Vienna University of Technology (TU Wien), A--1040 Vienna, Austria}
}
\affiliation{
\normalsize{
$^2$Department of Physics of Complex Systems,
Weizmann Institute of Science, Rehovot 761001, Israel}
}
\date{\today}

\begin{abstract}
The Casimir force between dielectric bodies is well--understood, but not the Casimir force inside a dielectric, in particular its renormalization. We develop and analyse a simple model for the Casimir forces inside a medium that is completely free of renormalization and show then how renormalization emerges. We consider a one--dimensional chain of point particles interacting with each other by scattering the zero--point fluctuations of the electromagnetic field confined to one dimension.  We develop a fast, efficient algorithm for calculating the forces on each particle and apply it to study the macroscopic limit of infinitely many, infinitely weak scatterers. The force density converges for piece--wise homogeneous media, but diverges in inhomogeneous media, which would cause instant collapse in theory. We argue that short--range counter forces in the medium prevent this collapse in reality. Their effect appears as the renormalization of the Casimir stress in dielectrics. Our simple model also allows us to derive an elementary analogue of the trace anomaly of quantum fields in curved space. 
\end{abstract}

\maketitle

\section{Introduction}

The Casimir effect \cite{Milonni,Milton,BKMM,Rodriguez,Buhmann,Forces,State} surely is one of the most fascinating and ``one of the least intuitive consequences of quantum electrodynamics'' \cite{Schwinger}. Yet it all began with a mundane, practical question \cite{LamoreauxPhysToday}: under which conditions are colloidal suspensions \cite{Israelachvili} --- such as liquid paint --- stable? A colloidal suspension is a fluid containing particles that repel each other on short distances by steric forces and attract each other on longer distances by van der Waals forces \cite{Hermann}. The balance of forces determines whether the suspension is stable \cite{Israelachvili}, for example, whether liquid paint stays liquid. 

Casimir was confronted with the following problem. Measurements indicated that the interparticle forces fall off like $r^{-7}$ with distance $r$, but London's theory \cite{London}  of static van der Waals forces predicted a $r^{-6}$ dependance. Overbeck suggested \cite{CasimirPolder} that retardation weakens the van der Waals force. Following this idea, Casimir and Polder \cite{CasimirPolder} deduced the $r^{-7}$ asymptotics from a full quantum--electrodynamical theory \cite{ShahmoonvdW}. Their theory \cite{CasimirPolder} was complicated and cumbersome, but their result was surprisingly simple and only depending on natural constants. Upon showing it to Bohr, he ``mumbled something about zero--point energy'' \cite{Letter}, which put Casimir on the track to discover the Casimir effect: \cite{Casimir} two planar perfect conductors in empty space attract each other by modifying the zero--point energy of the electromagnetic field. That energy is infinite, but the modifications are finite. 

Lifshitz \cite{Lifshitz} generalized the theory from perfect conductors to planar dielectrics separated by vacuum. Dzyaloshinskii and Pitaevskii joined in, replaced the vacuum by a homogeneous fluid in thermal equilibrium, and still managed to extend the theory to this case \cite{DLP}. Schwinger {\it et al.} confirmed their theory using a different starting point \cite{Schwinger}. The fluid gives rise to rich physical phenomena observed relatively recently: repulsive Casimir forces \cite{Milling,Meurk,Lee,Levitation} and stable Casimir equilibria \cite{CasimirEquilibrium}. Two planar dielectrics separated by a homogeneous fluid constitute a piece-wise homogeneous planar medium. Dzyaloshinskii and Pitaevskii thought \cite{DP} their theory applies to inhomogeneous dielectrics as well, but they were wrong \cite{SimpsonThesis}.

The problem is the renormalization required to extract the finite, physically valid force density from the infinite zero--point energy. It took modern computational tools to discover \cite{Simpson} that the renormalization procedure \cite{DP} of Dzyaloshinskii and Pitaevskii fails in general. Some partial solutions of the renormalization problem are known \cite{Renormalization,Itay2,Parashar,Li,Itai,Shayit,Vinas} but lack a sound physical justification. 

Here we address the problem of renormalization in inhomogeneous media with a theory inherently free of renormalization. We take as our starting point Casimir's initial problem of neutral particles in a fluid and simplify it further, inspired by the scattering approach to Casimir forces \cite{Lambrecht,Wirzba,Ingold}. We consider a one--dimensional chain of point scatterers interacting with each other by virtual electromagnetic waves confined to one dimension \cite{Fiedler} (by, {\it e.g.}, a waveguide \cite{Shahmoon}). Varying the scattering strengths of the particles we implement inhomogeneous dielectrics. The scatterers are immersed in a fluid of uniform refractive index, but by rescaling spatial coordinates we can replace the fluid by vacuum and calculate the van der Waals forces in the new coordinates as vacuum forces.   

The force on each scatterer is always finite. We find, however, that in the macroscopic limit of infinitely many, infinitely weak scatterers, the force density tends to infinity if they form an inhomogeneous medium. If, on the other hand, the scatterers are arranged to be piece--wise homogeneous, the force density converges to finite values, except at interfaces. Perturbations of homogeneity, however small they might be, would produce infinite forces in the macroscopic limit. Liquid paint would instantly solidify, unless it is absolutely homogenous, which does not happen in reality. Something is missing in the theory. What is it?

We follow and refine an argument by Pitaevskii \cite{Pitaevskii} that local restoring forces of the background fluid counteract the van der Waals forces on the scatterers. The net result is finite in the macroscopic limit. These counter forces are naturally occurring in colloidal suspensions. It is well--known \cite{Israelachvili} that they may stabilize homogeneous suspensions, here we argue that they also prevent a catastrophic collapse in inhomogeneous fluids. We identify how the counter forces are expressed in terms of macroscopic dielectric quantities. Mathematically, they appear as renormalizers, in agreement with previous renormalization procedures in inhomogeneous dielectrics \cite{Renormalization,Itay2,Parashar,Li,Itai,Shayit,Vinas} and in agreement with Lifshitz theory \cite{DLP}, but physically they are quite something else. Instead of seeing renormalization as a mathematical procedure to remove infinities, we regard it as the result of a physical process, the local equilibrium of forces. 

Inhomogeneities in fluids play a role in surface forces, and a better theory of van der Waals forces in inhomogeneous liquids may solve some of the mysteries there \cite{Sivan}. Realizing the limitations of Lifshitz theory may also help explaining the disagreement between theory and experiment in the Casimir force at long distances \cite{LifshitzDrude}. Finally, the renormalization of Casimir forces in inhomogeneous media is essential for understanding the quantum vacuum in curved space \cite{BD}. This is because curved space--time \cite{LL2} acts just like an inhomogeneous dielectric medium on electromagnetic fields \cite{Gordon,Plebanski,Schleich,LeoPhil}. We show how in our model the trace anomaly \cite{Itai,Wald} arises that plays the role of the cosmological constant \cite{Annals,Tkatchenko}. We thus have reasons to believe that our model, despite its simplicity, elucidates the essentials of some diverse physics, ranging from liquid paint to the cosmos. 

\section{The model}

\begin{figure}[h]
\begin{center}
\includegraphics[width=20pc]{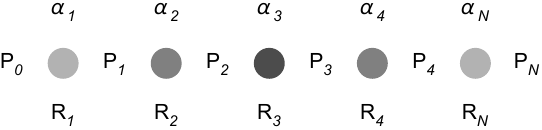}
\caption{
\small{The model: one--dimensional chain of $N$ point particles (here $N=5$) with varying polarizabilities $\alpha_i$ (varying shades of gray) interacting with each other by scattering electromagnetic waves in the vacuum state, freely propagating between them. The propagation is described by P matrices [Eq.~(\ref{eq:pmatrix})] and the scattering by R matrices [Eq.~(\ref{eq:R})].
}
\label{fig:chain}}
\end{center}
\end{figure}

\begin{figure}[h]
\begin{center}
\includegraphics[width=20pc]{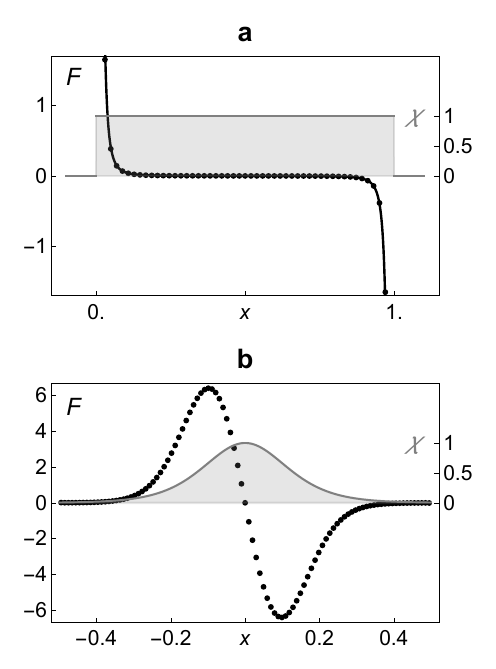}
\caption{
\small{van der Waals forces. {\bf a}: homogeneous block of $N=51$ particles (with $\chi$ illustrated by the gray filled curve) placed between $x_1=0$ and $x_{21}=1$. The dots show the numerically calculated van der Waals forces in units of $\hbar c$ and the solid black curve the analytic formula (\ref{eq:vdW}) in the macroscopic limit. They agree well for $N\ge 20$. The force peaks at the outer particles (not shown) and is attractive and antisymmetric; the net force is zero.  {\bf b}: $N=101$ particles with inhomogeneous profile [Eq.~(\ref{eq:sech2}) with the parameters $\chi_0=1$, $a=0.15$]. The force density diverges in the macroscopic limit ($N\rightarrow\infty$).
}
\label{fig:forces}}
\end{center}
\end{figure}

Consider a one--dimensional chain of $N$ point scatterers at positions $x_i$ with constant distance $\delta$ from each other (Fig.~\ref{fig:chain}). The electric polarizability $\alpha_i$ of each scatterer may vary such that they constitute a medium in the macroscopic limit $\delta\rightarrow 0$ with 
\begin{equation}
\chi(x) = \frac{\alpha_i}{\delta} 
\label{eq:chi}
\end{equation}
and $\chi$ kept finite. The polarizabilities may depend on frequency (establishing a dispersive medium). We will show in Sec.~III that $\chi$ represents the electric susceptibility \cite{Fiedler} and
\begin{equation}
n = \sqrt{1+\chi}
\label{eq:n}
\end{equation}
the refractive index. One could also assume a varying distance $\delta$ and constant polarizability $\alpha$ or both  $\delta$ and $\alpha$ varying, but here we take $\delta$ as constant for simplicity. The scatterers shall interact with each other by the vacuum forces \cite{Rodriguez,Buhmann,Forces,State} of the electromagnetic field confined to one--dimensional propagation. These are the van der Waals forces with both retardation and multiple scattering taken into account. In the following we work out what those forces are (Fig.~\ref{fig:forces}).

\subsection{Vacuum forces}

According to Hamilton's equations, the force on each scatterer is given by the negative derivative of the Hamiltonian with respect to the position $x_i$. For the Hamiltonian we take the zero--point energy of the electromagnetic field expressed as the integral of all vacuum energies $E/2$ times the density of states $D(E)$. Here $E$ represents the energy $E=\hbar\omega$ corresponding to the frequency $\omega$. The density of states is given by the Friedel--Krein--Lloyd formula \cite{Friedel,Krein,Lloyd,Faulkner} in terms of the scattering phase $\varphi$ as $D(E)=(2\pi)^{-1}\partial_E \varphi$. We integrate by parts and obtain for the van der Waals force
\begin{equation}
F_i=\frac{1}{4\pi} \int_0^\infty \partial_i \varphi \, dE \,,\quad E=\hbar\omega \,.
\label{eq:force}
\end{equation}
The scattering phase is the sum of all the phases of the eigenvalues of the classical scattering matrix $\rm{S}$. Since $\det\rm{S}$ is the product of all those eigenvalues we may write:
\begin{equation}
\varphi = -i\ln\,\det \rm{S} \,.
\label{eq:scatteringphase}
\end{equation}
With Eqs.~(\ref{eq:force}) and (\ref{eq:scatteringphase}) we have expressed the quantum force on each scatterer in terms of classical quantities. Quantum mechanics enters though $E=\hbar\omega$ and the fact that the zero--point energy is $E/2$.

\begin{figure}[h]
\begin{center}
\includegraphics[width=15pc]{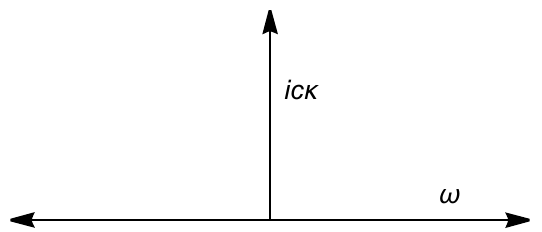}
\caption{
\small{Contour deformation. We represent the integrals $\int_0^{+\infty}$ and $-\int_{-\infty}^0$ over $\omega$ [Eq.~(\ref{eq:Fintegral})] by contours (shown as arrows) running to the right or the left, respectively, and move both contours to the positive imaginary axis ($\omega=ic\kappa$) where they coincide, which gives Eq.~(\ref{eq:forceformula}). 
}
\label{fig:deformation}}
\end{center}
\end{figure}

To proceed, we note that the scattering matrix is unitary, and that all unitary matrices can be expressed via the Cayley transform as 
\begin{equation}
{\rm S}=\frac{\mathbb{1}-i {\rm K}}{\mathbb{1}+i {\rm K}} \quad\mbox{ with Hermitian $\rm{K}$}.
\label{eq:Krepresentation}
\end{equation}
We extend ${\rm S}(E)$ to negative energies corresponding to negative frequencies. As the classical electromagnetic field is real, the negative--frequency component is the complex conjugate of the positive frequency component and runs backwards in time. Consequently,  ${\rm S}(-E)$ must be the inverse of  ${\rm S}(E)$. From this follows that
\begin{equation}
{\rm K}(-E)=-{\rm K}(E) \,.
\label{eq:Knegative}
\end{equation}
We thus obtain
\begin{equation}
F_i=\frac{i}{4\pi} \left(\int_0^{+\infty}-\int_{-\infty}^0\right) \frac{\partial_i\det(\mathbb{1}+i \rm{K})}{\det(\mathbb{1}+i \rm{K})} \,dE
\label{eq:Fintegral}
\end{equation}
and by deforming the integration contours on the complex plane (Fig.~\ref{fig:deformation}) to the positive imaginary axis, and expressing them as an integral over imaginary wave numbers $\kappa$ with
\begin{equation}
\omega = ic\kappa
\label{eq:kappa}
\end{equation}
and $c$ being the speed of light in vacuum, we get a rapidly converging formula for the force:
\begin{equation}
F_i=-\frac{\hbar c}{2\pi} \int_0^\infty  \frac{\partial_i\det(\mathbb{1}+i \rm{K})}{\det(\mathbb{1}+i \rm{K})} \,d\kappa \,.
\label{eq:forceformula}
\end{equation}
We are going to show (Sec.~IIC) that $\det(\mathbb{1}+i \rm{K})$ is real on the imaginary axis. Note that we have not performed any renormalization to calculate the force of the quantum vacuum on the point scatterers. 

\subsection{Transfer matrix} 

The only inconvenient feature of formula (\ref{eq:forceformula}) is the appearance of the scattering matrix (via $\rm{K}$). However, for one--dimensional propagation we may express $\det(\mathbb{1}+i \rm{K})$ in terms of the transfer matrix $\rm{T}$. The transfer matrix is much easier to calculate than the scattering matrix, because the total transfer matrix of the chain of scatterers is simply the product of the individual transfer matrices \cite{Transfer}. 

Consider real frequencies first. A right--moving wave $e^{ikx}$ on the left of the structure is transferred to the superposition $\alpha\, e^{ikx}+\beta\, e^{-ikx}$ with complex $\alpha$ and $\beta$ on the right (not to be confused with the polarizabilities $\alpha_i$) while a left--moving wave $e^{-ikx}$ on the left gets transferred to $\beta^*\, e^{ikx}+\alpha^*\, e^{-ikx}$ on the right. The transfer matrix has thus the structure \cite{Transfer}
\begin{equation}
{\rm T} = \begin{pmatrix}
\alpha & \beta^* \\
\beta & \alpha^*
\end{pmatrix}
\label{eq:T}
\end{equation}
with $\det \rm{T} = |\alpha|^2-|\beta|^2>0$ as the outgoing waves must not exceed the ingoing waves. The scattering matrix ${\rm S} $ describes how ingoing waves are turned into outgoing waves. The ingoing waves enter as $e^{ikx}$ on the left and $e^{-ikx}$ on the right, while the outgoing waves emerge as $e^{-ikx}$ on the left and $e^{ikx}$ on the right. One verifies that the coefficients of the outgoing and ingoing waves are related by the matrix 
\begin{equation}
{\rm S} = \frac{1}{\alpha^*} 
\begin{pmatrix}
\sqrt{|\alpha|^2-|\beta|^2} & \beta^* \\
-\beta & \sqrt{|\alpha|^2-|\beta|^2} 
\end{pmatrix} .
\label{eq:S}
\end{equation}
We see from Eq.~(\ref{eq:S}) that $\det S = \alpha/\alpha^*$ and infer from Eq.~(\ref{eq:T}) and $\det S =\det (\mathbb{1}+i{\rm K})^*/\det (\mathbb{1}+i{\rm K})$ the relation
\begin{equation}
\det (\mathbb{1}+i{\rm K})=T_{22} \,.
\label{eq:det}
\end{equation}
This relation, once established on the real axis of $E$, is extended by complex continuation to the imaginary axis (Fig.~\ref{fig:deformation}) where we need it.

Having made the connection between the transfer matrix and the force, we use the product rule of transfer matrices to express $\rm{T}$ in terms of the individual transfer matrices. Between the scatterers at positions $x_i$ the field propagates freely (Fig.~\ref{fig:chain}). The transfer matrices of free propagation are given by 
\begin{equation}
{\rm P}_i = \begin{pmatrix}
e^{-\kappa\delta_i} & 0 \\
0 & e^{\kappa\delta_i} 
\end{pmatrix}
\quad\mbox{with}\quad \delta_i=x_{i+1}-x_i
\label{eq:pmatrix}
\end{equation}
for purely imaginary wavenumbers where waves decay exponentially in propagation direction. For the sake of calculating the force, we keep the distance $\delta_i$ variable, as we need to differentiate with respect to the coordinates. The beginning and the end of the propagation region outside of the scatterers we put at the fixed points $x_0$ and $x_{N+1}$. The precise values of these points will not influence the force.  It is wise to break the total transfer matrix into two parts ${\rm A}_j$ and ${\rm B}_j$ describing the transfer from $x_0$ to $x_{j-1}$ and from $x_j$ to $x_{N+1}$, respectively. We denote the transfer matrices of the scatterers by ${\rm R}_i$ and obtain for the partial transfer matrices the recurrence relations
\begin{equation}
{\rm A}_{j+1}= {\rm P}_j\, {\rm R}_j\, {\rm A}_j
\quad\mbox{and}\quad
{\rm B}_{j-1} = {\rm B}_j \, {\rm R}_j \, {\rm P}_{j-1}
\label{eq:recurrence}
\end{equation}
with the initial values
\begin{equation}
{\rm A}_1= {\rm P}_0
\quad\mbox{and}\quad
{\rm B}_N = {\rm P}_N \,.
\label{eq:initial}
\end{equation}
With the help of the A and B matrices we get for the total transfer matrix:
\begin{equation}
{\rm T} = {\rm B}_j \, {\rm R}_j \, {\rm A}_j \,.
\label{eq:TAB}
\end{equation}
As $\partial_j =\partial_{\delta_{j-1}}-\partial_{\delta_j}$ we obtain from Eq.~(\ref{eq:pmatrix}) for the derivatives of the transfer matrix 
\begin{equation}
\partial_j{\rm T} = \kappa {\rm B}_j \left[\sigma_z, {\rm R}_j\right] {\rm A}_j
\label{eq:dT0}
\end{equation}
written in terms of the commutator with the Pauli matrix $\sigma_z={\rm diag}(1,-1)$. 

We have arrived at expressions for the force that are convenient for numerical calculations (Appendix A) and, as we shall see in Sec.~III, for deducing the force density in the macroscopic limit.

\subsection{Point scatterers}

It remains to specify the transfer matrices ${\rm R}_i$ of the point scatterers (Fig.~\ref{fig:chain}). We describe a point scatterer at position $x_i$ as a delta--function susceptibility in the Helmholtz equation of the electromagnetic field. We only consider one polarization component of the field and denote it by $\psi$ (for the other polarization we would get the same). For purely imaginary wavenumbers we thus have
\begin{equation}
\partial_x^2\psi = \kappa^2\left[1+\alpha_i\delta(x-x_i)\right]\psi \,.
\label{eq:helmpoint}
\end{equation}
The field to the left of $x_i$ we write as the superposition $a_-e^{-\kappa x}+a_+e^{\kappa x}$ while the field on the right is given by the (different) superposition $b_-e^{-\kappa x}+b_+e^{\kappa x}$. To satisfy Eq.~(\ref{eq:helmpoint}) the field $\psi$ must be continuous at $x_i$ while the derivative $\psi'$ is discontinuous and changes by $\kappa^2\alpha_i\psi(x_i)$. From this we obtain the connection between the coefficients, {\it i.e.}\ the transfer matrix: 
\begin{equation}
{\rm R}_i = \mathbb{1} - \frac{\kappa \alpha_i}{2}(\sigma_z+i\sigma_y) 
\label{eq:R}
\end{equation}
expressed here in terms of the Pauli matrices
\begin{equation}
\sigma_x = \begin{pmatrix}
0 & 1 \\
1 & 0
\end{pmatrix}
,\quad
\sigma_y = \begin{pmatrix}
0 & -i \\
i & 0
\end{pmatrix}
,\quad
\sigma_z = \begin{pmatrix}
1 & 0 \\
0 & -1
\end{pmatrix}
\label{eq:pauli}
\end{equation}
for mathematical convenience. Note that the ${\rm R}$ matrices are real, and so are  the ${\rm P}$ matrices (\ref{eq:pmatrix}) describing the free propagation between the scatterers. Therefore the total transfer matrix and its derivatives are real, too, and so is $\det(\mathbb{1}+i \rm{K})$ as we have stated in Sec.~IIA. From the commutation relations of the Pauli matrices (\ref{eq:pauli}) we get the relation $[\sigma_z,{\rm R}_i]=-\kappa\alpha_i\sigma_x$ and hence from Eq.~(\ref{eq:dT0}):
\begin{equation}
\partial_j{\rm T} = -\kappa^2 \alpha_j  {\rm B}_j \sigma_x {\rm A}_j \,.
\label{eq:dT}
\end{equation}
This completes the calculation of the van der Waals force on a chain of point particles using scattering theory and the transfer matrix. 

\section{Macroscopic limit}

In the macroscopic limit we assume that the inter--particle distance $\delta$ goes to zero with finite ratio (\ref{eq:chi}) between polarizability $\alpha_i$ and $\delta$. We translate the recurrence relations (\ref{eq:recurrence}) into differential equations and express the solution for the force density in terms of the macroscopic Green function. 

\subsection{Differential equations} 

Take the partial transfer matrices A and B as continuous functions of $x$ evaluated at the points $x_j$ of the scatterers. Substituting $\alpha_i=\chi\delta$ in Eq.~(\ref{eq:R}) and using the recurrence relations (\ref{eq:recurrence}) for the A and B matrices we linearize the differences ${\rm A}_{j+1}-{\rm A}_j$ and ${\rm B}_j-{\rm B}_{j-1}$ taking also into account the $\delta$--dependence in the propagation matrices (\ref{eq:pmatrix}). We obtain the differential equations
\begin{gather}
\partial_x {\rm A} = -\kappa {\rm V} {\rm A} 
\quad\mbox{and}\quad
\partial_x {\rm B} = \kappa {\rm B} {\rm V} 
\quad\mbox{with}
\nonumber\\
{\rm V} = \sigma_z + \frac{\chi}{2}(\sigma_z+i\sigma_y) 
\label{eq:diff}
\end{gather}
in terms of the Pauli matrices (\ref{eq:pauli}). We obtain from the initial values (\ref{eq:initial}) in the macroscopic limit (where ${\rm P}\rightarrow\mathbb{1}$) the initial conditions
\begin{equation}
{\rm A}(x_0)={\rm B}(x_{N+1}) = \mathbb{1} \,.
\label{eq:initialcond}
\end{equation}
According to Eqs.~(\ref{eq:forceformula}) and (\ref{eq:det}) the force is given in terms of the matrix element $T_{22}$ and its spatial derivative $T_{22}'$. Since ${\rm R}\rightarrow \mathbb{1}$ for $\delta\rightarrow 0$ we obtain from Eq.~(\ref{eq:TAB}) ${\rm T}={\rm BA}$, and so
\begin{equation}
T_{22}  = B_{21} A_{12} + B_{22} A_{22} \,.
\label{eq:T22}
\end{equation}
From Eq.~(\ref{eq:dT}) we get
\begin{equation}
T_{22}'  = -\kappa^2 \chi\delta \left(B_{21} A_{22} + B_{22} A_{12} \right) .
\label{eq:dT22}
\end{equation}
One verifies that for the relevant matrix elements the solutions of the differential equations (\ref{eq:diff}) have the structure
\begin{eqnarray}
A_{12}=(1-\kappa^{-1} \partial_x)\psi_+&,& A_{22}=(1+\kappa^{-1} \partial_x)\psi_+ \nonumber\\
B_{21}=(1+\kappa^{-1} \partial_x)\psi_- &,& B_{22}=(-1+\kappa^{-1} \partial_x)\psi_- \quad
\label{eq:structure}
\end{eqnarray}
where the $\psi_\pm$ satisfy the Helmholtz equation
\begin{equation}
\partial_x^2\psi = \kappa^2(1+\chi)\psi \,.
\label{eq:helmholtz}
\end{equation}
From the initial condition (\ref{eq:initialcond}) and the structure (\ref{eq:structure}) of the solution follows the asymptotics
\begin{equation}
\psi_\pm \propto e^{\pm\kappa x} \quad\mbox{for}\quad x\rightarrow\mp \infty \,.
\label{eq:asy}
\end{equation}
Taken all this together we obtain
\begin{equation}
\frac{T_{22}'}{T_{22}} = -\kappa^2\chi\delta \, \frac{(\psi_+\psi_-)'}{W}
\label{eq:Tratio}
\end{equation}
where $W$ denotes the Wronskian
\begin{equation}
W = \psi_+\psi_-'- \psi_-\psi_+' \,.
\label{eq:wronskian}
\end{equation}
Note that the Wronskian (\ref{eq:wronskian}) is constant in space as a consequence of the Helmholtz equation (\ref{eq:helmholtz}).

We have seen that the transfer across the scattering structure (and hence the force) is governed by the macroscopic Helmholtz equation (\ref{eq:helmholtz}). This shows that the $\chi$ defined in Eq.~(\ref{eq:chi}) is indeed the susceptibility and Eq.~(\ref{eq:n}) gives the refractive index. In our one--dimensional case \cite{Fiedler} the susceptibility depends linearly on the density $\delta^{-1}$. In three dimensions, this is normally only the case for dilute fluids. 

\subsection{Green function}

In the following we relate the force density in the macroscopic limit to the Green function. The Green function $g(x,x_0)$ describes the field propagation from a monochromatic point source of unit strength at position $x_0$ to the observation point $x$, and thus obeys the differential equation
\begin{equation}
(\partial_x^2 - \kappa^2n^2)g = \delta(x-x_0)
\label{eq:greeneq}
\end{equation}
with the asymptotics of a purely outgoing wave at infinity:
\begin{equation}
g \propto e^{\pm\kappa x} \quad\mbox{for}\quad x\rightarrow\mp \infty \,.
\label{eq:greena}
\end{equation}
Away from the point of emission ($x\neq x_0$) the Green function must be a solution of the Helmholtz equation (\ref{eq:helmholtz}). At the emission point $g$ must be continuous;
\begin{equation}
g(x_0+0,x_0) = g(x_0-0,x_0) 
\label{eq:x0a}
\end{equation}
while the derivative of $g$ needs to jump by $1$ in order to generate the delta function on the right--hand side of the Green equation (\ref{eq:greeneq}):
\begin{equation}
 \left.\partial_x g(x,x_0)\right|_{x=x_0+0} - \left.\partial_x g(x,x_0)\right|_{x=x_0-0} = 1 \,.
\label{eq:x0b}
\end{equation}
The well--known solution of these requirements is the expression
\begin{equation}
g(x,x_0) = 
	\frac{1}{W}\begin{cases}
		\psi_+(x)\, \psi_-(x_0) & :\quad x\le x_0 \,,\\
		\psi_+(x_0)\, \psi_-(x) & :\quad x\ge x_0 \,.
	\end{cases}
\label{eq:green}
\end{equation}

We turn to our calculation of the force in the macroscopic limit, and define the force density as 
\begin{equation}
f = \frac{F}{\delta} \,.
\end{equation}
We obtain from Eqs.~(\ref{eq:forceformula}), (\ref{eq:det}), (\ref{eq:Tratio}) and (\ref{eq:green}) the expressions
\begin{equation}
f = \frac{\hbar c}{2\pi} \int_0^\infty \widetilde{f} \, d\kappa
\,,\quad 
 \widetilde{f} = \kappa^2 \chi\, \partial_x g(x,x) \,.
 \label{eq:fresult}
\end{equation}
This is one of the central results of this paper. Similar to the Lifshitz theory \cite{DLP} of the Casimir force we have related the force density to the electromagnetic Green function, but in our case this is the force density {\it inside} dielectric media, not the force {\it between} dielectric bodies as in Lifshitz's case \cite{DLP}. Another important difference is the fact that our $g$ is the unrenormalized Green function --- at least for the time being. We shall discuss in Sec.~V how renormalization emerges. 

Figure \ref{fig:forces}a shows a simple example: the van der Waals forces inside a block of uniform material of length $a$ and refractive index $n_0$ surrounded by vacuum. We see that the force density peaks at the edges of the block, strongly pulling the material in due to van der Waals attraction, whereas near the middle of the block the van der Waals forces from the different parts of material nearly compensate each other. We also see the good agreement between the numerical calculation with the prediction (\ref{eq:fresult}) of the macroscopic theory, even for quite a small number of point scatterers ($\ge 50$).

\subsection{Asymptotics}

In order to understand whether and when formula (\ref{eq:fresult}) for the force density converges in the macroscopic limit, we apply the Madelung representation \cite{Madelung} for the fields $\psi_\pm$. The Helmholtz equation (\ref{eq:helmholtz}) plays the role of the Schr\"{o}dinger equation in optics, and Madelung has reduced the Schr\"{o}dinger dynamics to fluid mechanics \cite{Madelung}. Following Madelung \cite{Madelung} we make the  ansatz:
\begin{equation}
\psi_\pm = \frac{1}{\sqrt{2k}}\,\exp\left(\pm\int k\,dx\right)
\label{eq:madelung}
\end{equation}
where the Wronskian (\ref{eq:wronskian}) is exactly
\begin{equation}
W=-1 \,.
\label{eq:mw}
\end{equation}
From the Helmholtz equation (\ref{eq:helmholtz}) follows an equation for $k$ that is independent on the $\pm$ sign  in the exponent,
\begin{equation}
k^2 = n^2\kappa^2 -2\beta \,,
\label{eq:dispersion}
\end{equation}
and appears as a generalized dispersion relation with the extra term
\begin{equation}
\beta = \frac{\sqrt{k}}{2}\,\partial_x^2\, \frac{1}{\sqrt{k}} = - \frac{1}{4} \left(\frac{k'}{k}\right)' + \frac{1}{8} \left(\frac{k'}{k}\right)^2 .
\label{eq:beta}
\end{equation}
For the quantum--mechanical Schr\"{o}dinger equation, the $\beta$--term corresponds to Bohm's quantum potential \cite{Bohm}, in optics $\beta$ describes the scattering in an inhomogeneous medium. Mathematically, $\beta$ corresponds to the Schwarzian derivative \cite{AF} of the phase. 

For large frequencies ($\kappa\rightarrow\infty$) relation (\ref{eq:dispersion}) reduces to the ordinary dispersion relation $k^2\sim n^2\kappa^2$ of geometrical optics. We take the positive sign for $k$ such that the amplitude $(2k)^{-1/2}$ in the Madelung representation (\ref{eq:madelung}) is real. In subdominant order we take $k^2\sim[n\kappa - \beta/(n\kappa)]^2$ and have
\begin{equation}
k \sim n\kappa - \frac{\beta_0}{n\kappa} \,.
\label{eq:kasy}
\end{equation}
Here $\beta_0$ denotes the dominant contribution to $\beta$, {\it i.e.}\ Eq.~(\ref{eq:beta}) evaluated for $k=n\kappa$. We obtain
\begin{equation}
\beta_0 =  - \frac{1}{4} \left(\frac{n'}{n}\right)' + \frac{1}{8} \left(\frac{n'}{n}\right)^2 =  
- \frac{n''}{4n} + \frac{3n'^2}{8n^2} 
\label{eq:beta0}
\end{equation}
that does not explicitly depend on $\kappa$. The term $\beta_0$ quantifies the deviation from geometrical optics due to inhomogeneities in the medium, the amplitude of geometrical scattering. 

In the Madelung representation (\ref{eq:madelung}) we get for the Green function of Eq.~(\ref{eq:green}) evaluated at $x_0=x$ the simple expression $g(x,x) = - (2k)^{-1}$ and hence for the spectral force density of Eq.~(\ref{eq:fresult}):
\begin{equation}
\widetilde{f}=-\kappa^2 (n^2-1)\, \partial_x \frac{1}{2k} \,.
\label{eq:fresult1}
\end{equation}
For large frequencies (large imaginary wavenumbers $\kappa$) we obtain from Eq.~(\ref{eq:kasy}) the asymptotics:
\begin{equation}
\widetilde{f}\sim - \frac{n^2-1}{2} \,\partial_x \left(\frac{\kappa}{n}+\frac{\beta_0}{\kappa n^3}\right) .
\label{eq:fasy}
\end{equation}
For piece--wise homogeneous media (Figs.~\ref{fig:forces}a and \ref{fig:three}a) the derivative in Eq.~(\ref{eq:fasy}) vanishes, the integral (\ref{eq:fresult}) converges, and the force density is finite in the macroscopic limit. At dielectric interfaces (discontinuities of $n$) the force density does not converge, though. Moreover, any inhomoneity, however small, would produce an infinite force density. The suspension of scatterers would instantly and catastrophically collapse. This does not happen in reality. So which mechanism --- not considered in our model --- prevents the catastrophe? 

\section{Helmholtz force} 

It is wise to take a step back and relate our result (\ref{eq:fresult}) to the classical optical forces in dielectrics \cite{LL8}. We begin with a brief review on the Helmholtz force in three--dimensional electromagnetism. Then we turn to our one--dimensional model and identify the quantum forces and stresses there.

\subsection{Dielectric deformation}

Consider three--dimensional electromagnetism in a dielectric medium of electric permittivity $\varepsilon=n^2$ (and magnetic permeability $\mu=1$). Forces in the medium cause the dielectric to deform. Consider an infinitesimal deformation $\delta\bm{x}$. If we can write the variation of the free energy $F$ as the volume integral
\begin{equation}
\delta F = -\int \bm{f}\cdot\delta\bm{x}\, dV
\label{eq:fd}
\end{equation}
then $\bm{f}$ gives the force density, according to Hamilton's equations. The part of the free energy modified by the dielectric is the volume integral of the electric energy density $U=\frac{1}{2}\bm{E}\cdot\bm{D}$ with $\bm{D}=\varepsilon_0\varepsilon\bm{E}$ in SI units. Gauss' law $\nabla\cdot\bm{D}=\rho_{\rm ext}$ gives $\bm{D}$ in terms of the external electrical charges that are not modified by the deformation of the medium. We thus take $\bm{D}$ as fixed in the energy density and obtain $\delta U=-U \delta\varepsilon/\varepsilon$. Two terms contribute to $\delta\varepsilon$, the infinitesimal shift of the medium and the change in density $\rho$, as $\varepsilon$ may depend on density:
\begin{equation}
\delta\varepsilon = - \nabla\varepsilon\cdot\delta\bm{x} + \frac{\partial\varepsilon}{\partial\rho}\,\delta\rho
\label{eq:variation}
\end{equation}
Differential geometry \cite{Poisson} gives $\delta\rho=-\rho\nabla\cdot\delta\bm{x}$. Integrating by parts in the energy produces the structure of Eq.~(\ref{eq:fd}) with the force density \cite{LL8}
\begin{equation}
\bm{f} = - \frac{\nabla\varepsilon}{\varepsilon}\, \frac{\bm{E}\cdot \bm{D}}{2} + \nabla p_{\rm Ab}
\label{eq:hforce}
\end{equation}
and
\begin{equation}
p_{\rm Ab} = \frac{\partial\varepsilon}{\partial\rho}\,\rho\,\frac{\bm{E}\cdot \bm{D}}{2\varepsilon} \,.
\label{eq:Ab}
\end{equation}
The pressure $p_{\rm Ab}$ --- known as the Abraham pressure \cite{LL8} --- originates from the effect of the density variation on the dielectric: The total force (\ref{eq:hforce}) on the medium is called the Helmholtz force \cite{LL8}. For dilute media and in our one--dimensional case, $\varepsilon=1+\chi$ with $\chi=\alpha\rho/\varepsilon_0$ such that we get for the Helmholtz force:
\begin{equation}
\bm{f} = \chi \nabla \,\frac{\varepsilon_0 E^2}{2} = \rho \nabla \,\frac{\alpha E^2}{2} \,,
\label{eq:dipole}
\end{equation}
which is the dipole or gradient force familiar from optical tweezing \cite{Ashkin1,Ashkin2,Ashkin3,Neuman,Pesce}. 

\subsection{Quantum force} 

Let us relate our case of van der Waals forces on suspended particles to the Helmholtz force in dielectrics. For this we need to treat the electromagnetic field as a quantum field with the operator $\widehat{A}$ of the vector potential as fundamental quantity. The electric field strength $\widehat{E}$ is given by $-\partial_t\widehat{A}$ in Coulomb gauge. Consider the correlation functions \cite{Annals}
\begin{eqnarray}
K&=& \frac{\varepsilon_0 c}{2\hbar} \langle \widehat{A}_1 \widehat{A}_0 + \widehat{A}_0 \widehat{A}_1\rangle = \frac{\varepsilon_0 c}{\hbar} \,{\rm Re}\, \langle \widehat{A}_1 \widehat{A}_0 \rangle \,,
\label{eq:K}\\
\Gamma&=& \frac{\varepsilon_0 c}{2i\hbar} \langle \widehat{A}_1 \widehat{A}_0 - \widehat{A}_0 \widehat{A}_1\rangle = \frac{\varepsilon_0 c}{\hbar} \,{\rm Im}\, \langle \widehat{A}_1 \widehat{A}_0 \rangle 
\end{eqnarray}
where the indices refer to the positions $x_i$ and times $t_i$. Note that $\langle \widehat{A}_1 \widehat{A}_0 \rangle$ is an analytic function of the time difference $t=t_1-t_0$ if the field is in the vacuum state (or thermal) \cite{Annals}. For an analytic function the real and the imaginary part are connected \cite{AF}. Furthermore, $\Gamma$ corresponds to the difference between the retarded and the advanced Green function \cite{Annals}, the difference between outgoing and ingoing radiation: the dissipation. The correlation $K$ of the field fluctuations is thus related to the dissipation $\Gamma$, which is a form of the fluctuation--dissipation theorem \cite{Scheel}. 

In particular, in terms of our time--independent Green function $g$ and for real frequencies $\omega$ we have for the Fourier transform: \cite{Annals}
\begin{equation}
\widetilde{\Gamma} = \frac{g(\omega)-g(-\omega)}{2c} \,.
\label{eq:gammatilde}
\end{equation}
We see that $\widetilde{\Gamma}$ is an odd function of $\omega$. Consequently, we get for the inverse Fourier transform:
\begin{equation}
\Gamma(t) = \frac{1}{2\pi} \int_{-\infty}^{+\infty} \widetilde{\Gamma} e^{-i\omega t} \,d\omega 
= -\frac{i}{\pi} \int_0^\infty \widetilde{\Gamma} \sin\omega t \,d\omega \,.
\end{equation}
For finding the real part $K$ to the imaginary part $\Gamma$ we only need to replace $-\sin\omega t$ by $\cos\omega t$, because they represent the imaginary and real part of the analytic $e^{-i\omega t}$ of time evolution. \cite{Remark} Using Eq.~(\ref{eq:gammatilde}) and deforming the integration contours to the positive imaginary axis with wavenumbers $\kappa$ (Fig.~\ref{fig:deformation}) we obtain:
\begin{equation}
K(t) = - \frac{1}{\pi} \int_0^\infty g \cosh(c\kappa t)\,d\kappa \,.
\label{eq:Kformula}
\end{equation}
Strictly speaking, formula (\ref{eq:Kformula}) is only valid when $g$ falls off more rapidly than $\cosh(c\kappa t)$, which is the case outside of the light cone, $|x_1-x_0|>c|t|$, but only there we will need $K$. 

From the correlation function $K$ we can calculate the electromagnetic field correlations by differentiation. In particular, we get for the electric fields at equal times, but different positions:
\begin{equation}
\frac{\varepsilon_0}{2} \langle \widehat{E}_1 \widehat{E}_0 \rangle 
= \left.\frac{\hbar}{2c}\,\partial_{t_1}\partial_{t_0} K\right|_{t_1=t_0} 
= \frac{\hbar c}{2\pi} \int_0^\infty \kappa^2 g(x_1,x_0)\, d\kappa \,.
\label{eq:ee}
\end{equation}
Compare this with the result (\ref{eq:fresult}) for the force density in the macroscopic limit and Eq.~(\ref{eq:dipole}) for the Helmholtz force: the macroscopic van der Waals force is nothing else but the gradient force \cite{Ashkin1,Ashkin2,Ashkin3,Neuman,Pesce}. It is a peculiar gradient force, though. The force is not generated by external fields, but rather by the zero--point fluctuations of the field inside the medium. 

\subsection{Electromagnetic stresses}

Having established the force density, we are going to relate it now to the electric and magnetic stress tensors  \cite{LL8}
\begin{equation}
\sigma_E = \bm{E}\otimes\bm{D} - \frac{\bm{E}\cdot\bm{D}}{2} \,\mathbb{1}
\,,\quad
\sigma_M = \bm{H}\otimes\bm{B} - \frac{\bm{H}\cdot\bm{B}}{2} \,\mathbb{1}
\,.
\label{eq:sigma}
\end{equation}
From Maxwell's equations and the linear constitutive equations $\bm{D}=\varepsilon_0\varepsilon\bm{E}$ and $\bm{H}=\varepsilon_0 c^2\bm{B}$ follows the Abraham identity \cite{LL8} for stationary fields:
\begin{equation}
\nabla\cdot(\sigma_E + \sigma_M) = - \frac{\nabla\varepsilon}{\varepsilon} \, \frac{\bm{E}\cdot\bm{D}}{2} 
\label{eq:abraham}
\end{equation}
and from this we get for the Helmholtz force (\ref{eq:hforce}):
\begin{equation}
\bm{f} = \nabla\cdot(\sigma_E + \sigma_M) + \nabla p_{\rm Ab} \,.
\end{equation}
The force density is thus the divergence of the total electromagnetic stress plus the gradient of the Abraham pressure.

So far we considered the classical stress, but similarly to Sec.~IVB we can identify the quantum electromagnetic stresses as well. These stresses are the vacuum expectation values of expressions (\ref{eq:sigma}) for the field operators, symmetrized as in definition (\ref{eq:K}) of the correlation function $K$. From $\widehat{E}=-\partial_t \widehat{A}$ and $\widehat{B}=\partial_x \widehat{A}$ (in one dimension) we obtain: 
\begin{equation}
\widetilde{\sigma}_E = -n^2\kappa^2 g(x,x) 
\,,\quad
\widetilde{\sigma}_M = \left.\partial_x\partial_{x_0} g \right|_{x_0=x}
\label{eq:sigmas0}
\end{equation}
for the spectral stresses defined as in Eq.~(\ref{eq:fresult}). In terms of the Madelung representation [Eqs.~(\ref{eq:green}), (\ref{eq:madelung}) and (\ref{eq:mw})] we get:
\begin{equation}
\widetilde{\sigma}_E = \frac{n^2\kappa^2}{2k}
\,,\quad
\widetilde{\sigma}_M = \frac{k}{2} - \frac{k'^2}{8k^3} 
\label{eq:sigmas}
\end{equation}
with the wavenumber $k$ satisfying Eqs.~(\ref{eq:dispersion}) and (\ref{eq:beta}). We also express the quantum Abraham pressure (\ref{eq:Ab}) in terms of the Green function. In our one--dimensional case we have $\rho\,\partial\varepsilon/\partial\rho = \chi$ and so we obtain from Eq.~(\ref{eq:ee}) for the spectral Abraham pressure:
\begin{equation}
\widetilde{p}_{\rm Ab} = \chi \kappa^2 g(x,x) = -\frac{n^2-1}{n^2}\,\widetilde{\sigma}_E \,.
\label{eq:vacuumAb}
\end{equation}
One verifies that from Eqs.~(\ref{eq:dispersion}) and (\ref{eq:beta}) follows the Abraham identity for the quantum stresses (\ref{eq:sigmas}):
\begin{equation}
\partial_x (\widetilde{\sigma}_E+\widetilde{\sigma}_M) = \frac{2n'}{n}\,\widetilde{\sigma}_E \,.
\label{eq:quantumabraham}
\end{equation}

\begin{figure}[h]
\begin{center}
\includegraphics[width=20pc]{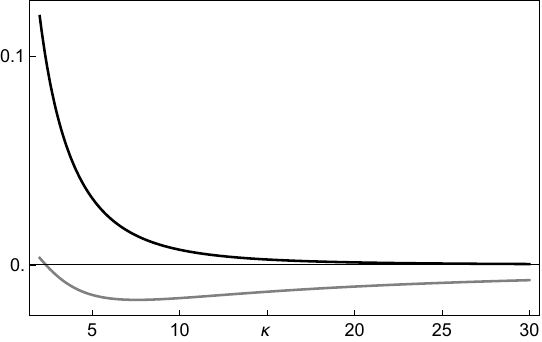}
\caption{
\small{Asymptotics of the stress. The figure illustrates the asymptotics of the spectral stress [Eq.~(\ref{eq:sigmatotal})] as a function of $\kappa$  for the susceptibility profile [Eq.~(\ref{eq:sech2})] displayed in Fig.~\ref{fig:forces}b (evaluated at $x=0.1$). Parameters as in Fig.~\ref{fig:forces}b. The gray curve shows $\widetilde{\sigma}-\kappa n$, the renormalization proposed by Dzyaloshinskii and Pitaevskii \cite{DP}. One sees that the curve gently decays (as $\kappa^{-1}$). The total renormalized stress is given by the integral over $\kappa$ and diverges. For the black curve the asymptotics (\ref{eq:asysigmatotal}) is subtracted; the curve decays as $\kappa^{-3}$ and this renormalization (Fig.~\ref{fig:renorm}) converges. 
}
\label{fig:asymptotics}}
\end{center}
\end{figure}

Formulas (\ref{eq:sigmas}) reveal the asymptotics of the stresses for large frequencies ($\kappa\rightarrow\infty$). Equation~(\ref{eq:kasy}) gives
\begin{equation}
\widetilde{\sigma}_E \sim \frac{\kappa n}{2}  + \frac{\beta_0}{2\kappa n}
\,,\quad
\widetilde{\sigma}_M  \sim \frac{\kappa n}{2}  - \frac{\beta_0}{2\kappa n} - \frac{n'^2}{8\kappa n^3} 
\label{eq:asysigma}
\end{equation}
such that we obtain for the total spectral stress,
\begin{equation}
\widetilde{\sigma}=\widetilde{\sigma}_E+\widetilde{\sigma}_M \,,
\label{eq:sigmatotal}
\end{equation}
the asymptotics:
\begin{equation}
\widetilde{\sigma} \sim \kappa n  - \frac{n'^2}{8\kappa n^3} \,.
\label{eq:asysigmatotal}
\end{equation}
Like in the three--dimensional planar case \cite{Renormalization} the asymptotics of the total stress depends only on the local dielectric functions and their first derivatives, in our case on the refractive index $n$ and its derivative $n'$ (Fig.~\ref{fig:asymptotics}). We also see that the integrated stresses diverge even in the case of piece--wise homogeneous media when the force density (\ref{eq:fresult}) converges.

\section{Renormalization} 

Following the scattering approach to Casimir forces \cite{Lambrecht,Wirzba,Ingold} we have developed a theory that directly gives the actual forces on point scatterers, without the need of renormalization. We found that the density of the van der Waals force and the electromagnetic stresses diverge in the macroscopic limit, in direct contradiction to the empirical fact that some colloidal suspensions are stable --- that liquid paint may remain liquid. Some additional feature must prevent this catastrophic collapse pure van der Waals forces would cause. 

\subsection{Argument}

Pitaevskii \cite{Pitaevskii} argued that the pressure of the fluid naturally counteracts the van der Waals forces. These counter forces depend on the density and are typically much stronger than the van der Waals forces, such that a small variation in density is sufficient to reach a local mechanical equilibrium. As this small density variation hardly affects the van der Waals force, we ignore it in calculating the force, but we do use the fact that the counter pressure it causes partly compensates the van der Waals force. This counter pressure appears as the renormalization of the Casimir force. 

Which contributions to the van der Waals force are compensated? Pitaevskii \cite{Pitaevskii} points out the following. As the counter pressure reacts to density variations, it immediately takes out the part of the quantum force that is also generated by density variations (Sec.~IVA): the Abraham pressure [Eq.~(\ref{eq:Ab})]. What remains is the divergence of the  electromagnetic stress in the material \cite{Pita,Burger,Burger2}. But, as we argue here, this is not all. The counter pressure can also react to local variations of the electromagnetic stresses, establishing local mechanical equilibria everywhere.

We need to define what we mean by local. The electromagnetic stresses are created by zero--point fluctuations of the field in inhomogeneous media. We may characterize the length scale $\ell$ of the inhomogeneity by $\ell^{-1}=|n'/n|$. Only frequencies with wavelengths $\lambda\ll\ell$ will contribute to the local stresses. As $\lambda$ is inversely proportional to $n$ we obtain the criterion
\begin{equation}
|\nabla\lambda|\ll 1 
\quad\mbox{with}\quad \lambda = \frac{2\pi}{\kappa n} \,,
\label{eq:criterion}
\end{equation}
which is the well--known condition for the validity of geometrical optics or of the semiclassical approximation in quantum mechanics \cite{LL3}. 

We can only define local stresses for fields consistent with geometrical optics. The field at zero frequency will never satisfy this condition, so whenever zero--frequency features are important, the renormalization procedure is bound to fail. This is the case for thermal Casimir forces dominated by the zero Matsubara frequency \cite{LL7}, which might explain why Lifshitz theory \cite{DLP} does not agree with experiments there \cite{LifshitzDrude}. 

\subsection{Local stresses}

Next we assume criterion (\ref{eq:criterion}) and work out the local electromagnetic stresses. According to Eq.~(\ref{eq:sigmas0}) the stresses are generated by the Green function. We must define a new Green function that captures their local contributions. This $g_0$ we only need in the vicinity of the point of emission $x_0$ and there we assume \cite{Renormalization} that it describes a purely outgoing wave unaffected by the refractive index profile away from $x_0$. Here is the argument for $g_0$:

Green functions must satisfy two conditions: the propagation equation (\ref{eq:greeneq}) and the boundary conditions. The total Green function $g$ is required by the boundary condition (\ref{eq:greena}) to be purely outgoing at $\pm\infty$. 
For $g_0$ we require that it is purely outgoing in an infinitesimal interval around $x_0$ where outgoing we define in the sense of geometrical optics. To the left or right of $x_0$ it should propagate like the $\psi_-$ or $\psi_+$ waves in the Madelung representation (\ref{eq:madelung}) with $k=n\kappa$. They obey the condition $\mathbb{D}_\pm \psi_\pm=0$ with the differential operator
\begin{equation}
\mathbb{D}_\pm = \partial_x \pm \kappa n + \frac{n'}{2n} \,.
\label{eq:dpm}
\end{equation}
The $ \partial_x \pm \kappa n$ term captures the propagation of the phase, while the $n'/(2n)$ term takes care of the amplitude variation in geometrical optics. We thus encode the boundary condition as
\begin{equation}
\mathbb{D}_\pm g_0=0 
\quad\mbox{for}\quad x=x_0\pm 0 \,.
\label{eq:out}
\end{equation}

Being a Green function, $g_0$ must solve the propagation equation (\ref{eq:greeneq}). For $x\neq x_0$ it is therefore a solution of the Helmholtz equation (\ref{eq:helmholtz}), {\it i.e.}\ a superposition of the fundamental system $\psi_\pm$ of wave functions:
\begin{equation}
g_0 = 
	\begin{cases}
		c_{+-}\psi_+(x)+c_{--}\psi_-(x) & :\quad x<x_0 \,,\\
		c_{++}\psi_+(x)+c_{-+}\psi_-(x) & :\quad x>x_0 \,.
	\end{cases}
\label{eq:super}
\end{equation}
Note that the coefficients depend on $x_0$. At the point of emission, $x=x_0$, the Green function $g_0$ must obey the continuity condition (\ref{eq:x0a}) and the discontinuity (\ref{eq:x0b}) of the derivative. The four conditions of Eqs. (\ref{eq:x0a}-\ref{eq:x0b}) and (\ref{eq:dpm}-\ref{eq:out}) determine the four coefficients of $g_0$ uniquely. We obtain 
\begin{equation}
c_{\pm a}=\left.\pm\frac{\mathbb{D}_a\psi_\mp}{2\kappa n} \right|_{x_0} \quad\mbox{for}\quad a=\pm \,.
\label{eq:coeff}
\end{equation}
With these coefficients for $g_0$ we calculate the spectral stresses (\ref{eq:sigmas0}) using the Helmholtz equation (\ref{eq:helmholtz}) for expressing $\psi_\pm''$ as $\kappa^2n^2\psi_\pm$. We obtain the exact result
\begin{equation}
\widetilde{\sigma}_{E0} = \frac{\kappa n}{2} 
\,,\quad
\widetilde{\sigma}_{M0} = \frac{\kappa n}{2}  + \partial_x \frac{n'}{4\kappa n^2} \,.
\label{eq:sigma0}
\end{equation}
We see that $\widetilde{\sigma}_{E0}$ and $\widetilde{\sigma}_{M0}$ depend only on the local refractive index $n$ and its first and second derivative. The local Green function $g_0$ does indeed capture local contributions to the electromagnetic stresses. Appendix B compares $g_0$ with the previous renormalizer \cite{Itai} and finds good agreement. 

\subsection{Anomaly}

We have seen that Eq.~(\ref{eq:sigma0}) describes local stresses, but as Eq.~(\ref{eq:asysigma}) indicates, they do not completely account for the asymptotics of the full quantum electromagnetic stresses at large frequencies ($\kappa\rightarrow\infty$). We use Eq.~(\ref{eq:beta0}) and write the full asymptotics (\ref{eq:asysigma}) in terms of the local stresses (\ref{eq:sigma0}) as
\begin{equation}
\widetilde{\sigma}_E \sim \widetilde{\sigma}_{E0} + \frac{\beta_0}{2\kappa n}
\,,\quad
\widetilde{\sigma}_M  \sim \widetilde{\sigma}_{M0} + \frac{\beta_0}{2\kappa n} \,.
\label{eq:asystresses}
\end{equation}
We see that in both the electric and the magnetic stress there is a characteristic difference by $\beta_0/(2\kappa n)$. For dispersive media $\beta_0$ will decay sufficiently fast such that the integral of the difference converges, but there is another, deeper problem: the stresses (\ref{eq:sigma0}) do not satisfy the Abraham identity (\ref{eq:quantumabraham}). This is because the full stresses do, and so do the asymptotic stresses (\ref{eq:asystresses}), but not the terms $\beta_0/(2\kappa n)$. Now, the Abraham identity describes momentum conservation and it is a consequence of Maxwell's equations. So why is it violated here? 

In our one--dimensional case, Maxwell's equations reduce to the Helmholtz equation. One can derive the Abraham identity (\ref{eq:quantumabraham}) for the stresses (\ref{eq:sigmas0}) from the Helmholtz equation for the Green function [Eq.~(\ref{eq:greeneq}) at $x\neq x_0$] and the reciprocity of the Green function:
\begin{equation}
g(x,x_0) = g(x_0,x) \,.
\label{eq:reciprocity}
\end{equation}
The reciprocity appears in the structure (\ref{eq:green}) of the Green function, and it stems from the boundary conditions (\ref{eq:greena}). They state that the field is purely outgoing at infinity. Therefore, the only source of radiation is the emission at point $x_0$. If there are no other sources from infinity we can swap the positions of emitter and receiver and get the same. The local Green function, on the other hand,  does satisfy the Helmholtz equation as well, but not the reciprocity, because in order to sustain its boundary conditions (\ref{eq:dpm}) and (\ref{eq:out}) sources at $\pm\infty$ are required. Another, direct argument is the following. For $g_0$ the boundary conditions (\ref{eq:dpm}-\ref{eq:out}) depend on the dielectric environment at the point of emission $x_0$. So the emission of radiation depends on $n(x_0)$ and $n'(x_0)$, but not the reception, which breaks the reciprocity (\ref{eq:reciprocity}). 

In order to restore momentum conservation [the Abraham identity (\ref{eq:quantumabraham})] we would need to add $\beta_0/(2\kappa n)$ to the spectral stresses $\widetilde{\sigma}_{E0}$ and $\widetilde{\sigma}_{M0}$. The extra term is called an anomaly \cite{Wald}. Wald \cite{Wald} predicted it first for quantum fields in curved space \cite{BD}. He derived a trace anomaly from the lack of reciprocity of the renormalizer if there is space--time curvature. Inhomogeneous dielectric media correspond to space--time geometries and vice versa \cite{Gordon,Plebanski,Schleich,LeoPhil}. So it is perhaps not surprising that we obtain an anomaly as well. Note however that the van der Waals anomaly is quantitatevely different from Wald's  anomaly \cite{Wald}. It was first deduced \cite{Itai} for three--dimensional media with spherical symmetry; here we have derived it for the simple one--dimensional case as well. 

For quantum fields in dielectric media, the anomaly describes the recoil pressure in the material \cite{LondonProceedings}. The picture is this: Each infinitesimal cell of the material is emitting and receiving virtual electromagnetic waves described by the Green function $g$, which creates the electromagnetic stresses. The local contributions to the stresses depend on the local Green function $g_0$, but for $g_0$ emission and reception are no longer interchangeable. The lack of reciprocity creates an imbalance in the recoil of emission and reception that appears as an additional pressure: the anomaly. 

Equations~(\ref{eq:sigma0}) derived from the local Green function plus the anomaly capture the divergent contributions to the electromagnetic stresses. As they are local the counter pressure of the suspension should be able to compensate for them. What remains are the effective electromagnetic stresses:
\begin{equation}
\widetilde{\sigma}_F^{\rm eff} = \widetilde{\sigma}_F - \widetilde{\sigma}_{F0} - \frac{\beta_0}{2\kappa n}
\quad\mbox{for}\quad F =E,M 
\label{eq:eff}
\end{equation}
with $\widetilde{\sigma}_{F0}$ given by Eq.~(\ref{eq:sigma0}) and $\beta_0$ by Eq.~(\ref{eq:beta0}).  We have thus deduced the renormalizing counter stresses and the anomaly from condition (\ref{eq:out}). Note that we should introduce a cutoff at low frequencies for being consistent with the criterion (\ref{eq:criterion}) for locality. In three dimensions \cite{Itai}  such a cutoff would not be necessary as the density of states goes with $\kappa^2$, which sufficiently suppresses the low--frequency contribution. There the criterion (\ref{eq:criterion}) only matters if the zero frequency becomes singled out and dominant, as in the thermal Casimir effect \cite{LifshitzDrude}.

\begin{figure}[h]
\begin{center}
\includegraphics[width=20pc]{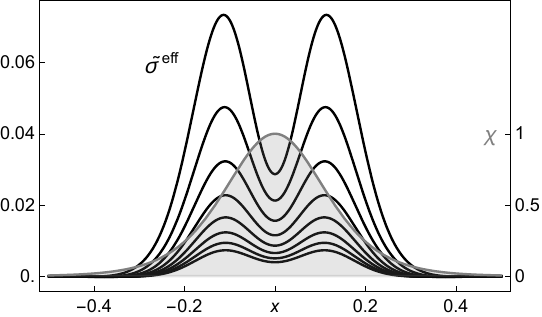}
\caption{
\small{Renormalized stress. The figure shows the renormalized total spectral stress $\widetilde{\sigma}^{\rm eff}=\widetilde{\sigma}_E^{\rm eff}+\widetilde{\sigma}_M^{\rm eff}$ with the effective stresses given by Eq.~(\ref{eq:eff}) for the susceptibility profile $\chi$ (gray filled curve) of Eq.~(\ref{eq:sech2}) for several values of $\kappa$ (from top to bottom: $\kappa$ from 3 to 10 in steps of 1). Figure~\ref{fig:forces}b shows the bare van der Waals forces, whereas here we see the effect of local counter forces (for the same parameters). The effective stress resembles the box--like profile [Eq.~(\ref{eq:stresses})] for the stress in a homogeneous layer, apart from swings at the edges. As the stress rises the force is attractive (whereas in Fig.~\ref{fig:three}b the stress drops and the Casimir force is repulsive). 
}
\label{fig:renorm}}
\end{center}
\end{figure}

\section{Examples}

Let us consider two examples to illustrate our ideas: first a piece--wise homogeneous dielectric of three layers and then a genuinely inhomogeneous medium with sech--squared profile of the susceptibility. The three layers constitute the classic case \cite{DLP} of Casimir forces. We will revisit this case and see in detail how the renormalization in dielectrics emerges there. 

\subsection{Three layers}

Consider a homogenous dielectric layer of thickness $a$ and refractive index $n_2$ sandwiched between two infinitely thick layers of homogeneous media with $n_1$ on the left and $n_3$ on the right. We put the coordinates such that the left interface lies at $x=0$ and the right interface at $x=a$. For calculating the force density according to Eq.~(\ref{eq:fresult}) we need the Green function at equal positions. In the central layer we have (Appendix C):
\begin{eqnarray}
g(x,x) &=& -\frac{1}{2n_2\kappa} - \frac{\varrho_l\, e^{-2n_2\kappa x} + \varrho_r\, e^{2n_2\kappa(x-a)}}{2n_2\kappa\,(1-\varrho_l\varrho_r\,e^{-2n_2\kappa a})} 
\nonumber\\
& & - \frac{1}{n_2\kappa}\, \frac{1}{(\varrho_l\varrho_r)^{-1} e^{2n_2\kappa a} - 1} 
\label{eq:gmiddle}
\end{eqnarray}
in terms of the Fresnel reflection coefficients 
\begin{equation}
\varrho_l = \frac{n_2-n_1}{n_2+n_1}  \quad\mbox{and}\quad \varrho_r = \frac{n_2-n_3}{n_2+n_3}\,,
\label{eq:fresnel}
\end{equation}
whereas in the left layer we obtain (Appendix C):
\begin{equation}
g(x,x)=-\frac{1}{2n_1\kappa}  + \frac{1}{2n_1\kappa} \,\frac{\varrho_l-\varrho_r\,e^{-2n_2\kappa a}}{1-\varrho_l\varrho_r\,e^{-2n_2\kappa a}}  \, e^{2n_1\kappa x} \,.
\label{eq:gleft}
\end{equation}
For the Green function in the right layer we only need to replace in Eq.~(\ref{eq:gleft}) the index $n_1$ by $n_3$ and the term $e^{2n_1\kappa x}$ by $e^{2n_3\kappa (a-x)}$ and to exchange $\varrho_l$ and $\varrho_r$. We now have all expressions ready for calculating the force density (\ref{eq:fresult}). 

\begin{figure}[h]
\begin{center}
\includegraphics[width=20pc]{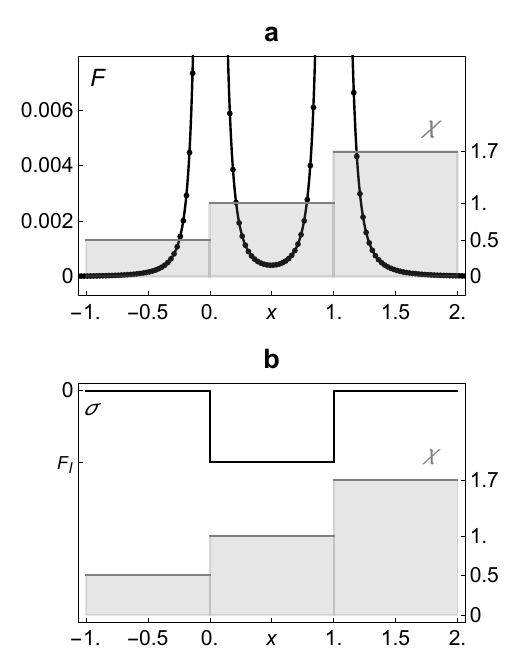}
\caption{
\small{Three layers. {\bf a}: The piece--wise homogeneous susceptibility profile $\chi$ (filled steps) generates the van der Waals forces $F$ (dots) on each point particle (in units of $\hbar c$). The solid curve shows the macroscopic limit of the forces [Eq.~(\ref{eq:vdW})]. The forces are attractive towards regions of higher susceptibility and peak at the interfaces (see also Fig.~\ref{fig:forces}a). {\bf b}: Counter forces in the material remove most of the van der Waals forces and produce the stress $\sigma$ that is zero in the outer layers and reaches the Lifshitz value [Eq.~(\ref{eq:lifshitz})] in the central layer. The discontinuities of the stress generate the Casimir--Lifshitz force at the interfaces [Eq.~(\ref{eq:lifshitzf})] --- a repulsive force here. 
}
\label{fig:three}}
\end{center}
\end{figure}

Assume that the refractive indices do not depend on frequency (no dispersion). In this case we can analytically perform the integration in formula (\ref{eq:fresult}) in terms of the Lerch transcendent [Sec.~1.11 of Ref. \onlinecite{Erdelyi}]:
\begin{equation}
\Phi(z,s,x)=\sum_{m=0}^\infty \frac{z^m}{(m+x)^s}\,,
\end{equation}
because $\Phi$ has the integral representation 1.11(3) of Ref. \onlinecite{Erdelyi}:
\begin{equation}
\Phi(z,s,x)=\frac{1}{\Gamma(s)} \int_0^\infty \frac{t^{s-1}e^{-x t}}{1-z e^{-t}} \,dt \,.
\end{equation}
We obtain in the three regions
\begin{eqnarray}
f_1 &=& \frac{\hbar c}{\pi} \frac{n_1^2-1}{(2n_2 a)^3}\left[\varrho_l\,\Psi\Big(-\frac{n_1 x}{n_2 a}\Big)-\varrho_r\,\Psi\Big(1-\frac{n_1 x}{n_2 a}\Big) \right] ,
\nonumber\\
f_2&=& \frac{\hbar c}{\pi} \frac{n_2^2-1}{(2n_2 a)^3}\left[\varrho_l\,\Psi\Big(\frac{x}{a}\Big)-\varrho_r\,\Psi\Big(1-\frac{x}{a}\Big) \right] ,
\label{eq:vdW}\\
f_3 &=& \frac{\hbar c}{\pi} \frac{n_3^2-1}{(2n_2 a)^3}\left[\varrho_l\,\Psi\Big(1-\frac{n_3 x_a}{n_2 a}\Big)-\varrho_r\,\Psi\Big(-\frac{n_3x_a}{n_2a}\Big) \right] 
\nonumber
\end{eqnarray}
with $x_a =a-x$ while $\Psi(\xi)$ abbreviates
\begin{equation}
\Psi(\xi)= \Phi(\varrho_l\varrho_r,3,\xi) \sim \xi^{-3} \quad\mbox{for}\quad \xi\rightarrow 0 \,.
\end{equation}
The force density is sharply peaked at the dielectric interfaces where the pull of the van der Waals forces is strongest (Figs.~\ref{fig:forces}a and \ref{fig:three}a). However, since $\chi$ is spatially constant, these forces [Eq.~(\ref{eq:fresult})] are entirely given by the Abraham pressure [Eq.~(\ref{eq:vacuumAb})]. We have argued in Sec.~VA that counter forces in the material completely compensate this pressure. 

What is left are the stress differences at the interfaces. From the Green equation (\ref{eq:greeneq}) follows that $g$ is some superposition of $e^{\pm n \kappa x}$ in each of the three layers and for $x\neq x_0$. Reciprocity [Eq.~(\ref{eq:reciprocity})] implies that $g(x,x_0)$ consists of terms proportional to $e^{\pm n \kappa (x-x_0)}$ and $e^{\pm n \kappa (x+x_0)}$. The $e^{\pm n \kappa (x+x_0)}$ terms have produced the position--dependent parts in $g(x,x)$ and hence the Abraham pressure. We see from Eq.~(\ref{eq:sigmas0}) that these terms cancel each other in the sum (\ref{eq:sigmatotal}) of the electromagnetic stresses, whereas the $e^{\pm n \kappa (x-x_0)}$ terms generate equal electric and magnetic stresses. In the limit $x_0\rightarrow x$ these are the position--independent parts of Eqs.~(\ref{eq:gmiddle}) and (\ref{eq:gleft}). For them, twice the electric stress $-n^2\kappa^2g$ gives the total stress:
\begin{gather}
\widetilde{\sigma}_1 = n_1\kappa  \,,\quad \widetilde{\sigma}_3 = n_3\kappa\,,
\nonumber\\
\widetilde{\sigma}_2 = n_2\kappa +  \frac{2n_2\kappa}{(\varrho_l\varrho_r)^{-1} e^{2n_2\kappa a} - 1} \,.
\label{eq:stresses}
\end{gather}
The stress is constant in each of the dielectric layers, but discontinuous at the interfaces, generating there the localized force densities 
\begin{equation}
f = F_l\,\delta(x) + F_r\,\delta(x-a) \,.
\label{eq:lifshitzf}
\end{equation}
The dominant contributions to the stress (\ref{eq:stresses}) are the $n\kappa$ terms in agreement with the asymptotics (\ref{eq:asysigma}). But these are entirely local terms that only depend on the local refractive index. We argued that they are also compensated by the local counter pressure. What remains are the non--local contributions that depend on the distance $a$ and the reflections coefficients, the Casimir--Lifshitz forces \cite{DLP}:
\begin{equation}
F_l = -F_r = \frac{\hbar c}{\pi} \int_0^\infty \frac{n_2\kappa\,d\kappa}{(\varrho_l\varrho_r)^{-1} e^{2n_2\kappa a} - 1} \,.
\label{eq:lifshitz}
\end{equation}
The van der Waals forces of Eq.~(\ref{eq:vdW}) are attractive towards regions of higher refractive index and peak at the dielectric interfaces (Figs.~\ref{fig:forces}a and \ref{fig:three}a). The counter forces in the material compensate them completely, except at the interfaces. The resulting Casimir--Lifshitz force of Eq.~(\ref{eq:lifshitz}) may be attractive or repulsive, depending on the sign of the product $\varrho_l\varrho_r$ of the Fresnel reflection coefficients (\ref{eq:fresnel}). For $n_1<n_2<n_3$ or $n_1>n_2>n_3$ the force is repulsive \cite{DLP} (Fig.~\ref{fig:three}b). One of the layers may even levitate on  quantum fluctuations \cite{Levitation,CasimirEquilibrium}.

For the classic case of three homogeneous layers --- and from our perspective on renormalization --- we have derived Eq.~(\ref{eq:lifshitz}): Lifshitz' formula.

\subsection{Inhomogeneous profile}

\begin{figure}[h]
\begin{center}
\includegraphics[width=20pc]{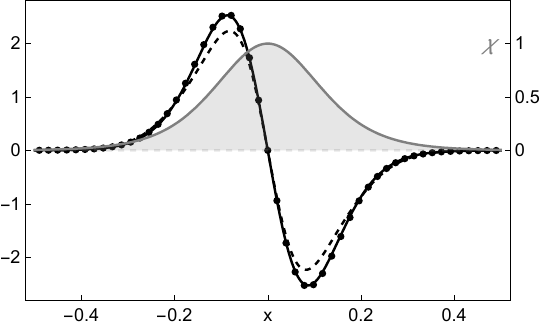}
\caption{
\small{Spectral forces. Comparison of the spectral force (dots) $-\partial_j \ln T_{22}$ [the integrand of the total force (\ref{eq:forceformula1})] with the macroscopic limit  (solid curve) [Eq.~(\ref{eq:fresult})] and the asymptotics (dashed curve) [Eq.~(\ref{eq:fasy})] for the susceptibility profile $\chi$ of Eq.~(\ref{eq:sech2}) (gray filled curve). Parameters as in Fig.~\ref{fig:forces}b, $\kappa=8.0$. The curves eventually merge for large $\kappa$, but are distinct yet at the moderate $\kappa$ used, for better visibility.
}
\label{fig:spectral}}
\end{center}
\end{figure}

Let us now turn to an inhomogeneous medium where standard Lifshitz renormalization fails \cite{Simpson}. We take an example where we can still calculate the Green function analytically, and compare the resulting forces and stresses with the asymptotics for large wavenumbers and with numerical results for a finite number of particles. Consider the susceptibility 
\begin{equation}
\chi = \chi_0 \sech^2(x/a)
\label{eq:sech2}
\end{equation}
as a function of the spatial coordinate $x$ in units of the characteristic scale $a$ of the profile. The Helmholtz equation (\ref{eq:helmholtz}) has the solution
\begin{equation}
\psi_+ = e^{\kappa x} {}_2 F_1(\nu_-,\nu_+,\kappa+1, \zeta) 
\,,\quad 
\zeta = \frac{1+\tanh(x/a)}{2}
\label{eq:psiplus}
\end{equation}
with the parameters
\begin{equation}
\nu_\pm = \frac{1}{2}\left(1\pm\sqrt{1-4\chi_0\,a^2\kappa^2}\right)
\label{eq:alphapm}
\end{equation}
in terms of Gauss' hypergeometric function \cite{Erdelyi} ${}_2 F_1$ as $\zeta$ reduces Eqs.~(\ref{eq:helmholtz})  and (\ref{eq:sech2}) to the hypergeometric differential equation \cite{Erdelyi}. Note that the $\nu_\pm$ are complex for $\kappa>1/(2a\sqrt{\chi_0})$ where $\nu_\mp$ becomes the complex conjugate of $\nu_\pm$. However, as the hypergeometric series \cite{Erdelyi} goes equally with $\nu_-$ and $\nu_+$ all the terms in the series are real and so Eq.~(\ref{eq:psiplus}) always describes a real function. For $x\rightarrow -\infty$ the variable $\zeta$ tends to zero. Since all hypergeometric functions are unity at zero argument the wave $\psi_+$ approaches $e^{\kappa x}$ and so satisfies the boundary condition (\ref{eq:asy}) as required. For constructing the second solution $\psi_-$ of the wave equation and of the opposite asymptotics ($\psi_-\sim e^{-\kappa x}$ for $x\rightarrow+\infty$) we take advantage of the mirror symmetry of the profile (\ref{eq:sech2}) and put
\begin{equation}
\psi_-(x) = \psi_+(-x) \,.
\label{eq:psiminus}
\end{equation}
For the Green function (\ref{eq:green}) we need the Wronskian (\ref{eq:wronskian}). The Wronskian is constant throughout $x$ and so we may calculate it where the calculation is most convenient, for $x\rightarrow+\infty$ where $\psi_-\sim e^{-\kappa x}$ and 
\begin{eqnarray}
\psi_+ &\sim& \frac{\Gamma(-\kappa a)\,\Gamma(-\kappa a-1)}{\Gamma(\kappa a+\nu_+)\,\Gamma(\kappa a+\nu_-)}\, e^{-\kappa x} 
\nonumber\\
&& + \frac{\Gamma(\kappa a)\,\Gamma(\kappa a+1)}{\Gamma(\kappa a+\nu_+)\,\Gamma(\kappa a+\nu_-)}\, e^{\kappa x} 
\label{asy}
\end{eqnarray}
according to Eq.~e7 of Ref. \onlinecite{LL3}. Here $\Gamma$ denotes the Gamma function \cite{Erdelyi} with $z \Gamma(z)=\Gamma(z+1)$. We thus obtain
\begin{equation}
W = -\frac{2[\Gamma(\kappa a+1)]^2}{a\Gamma(\kappa a+\nu_+)\,\Gamma(\kappa a+\nu_-)}\,.
\end{equation}
Figure~\ref{fig:forces}b shows the bare van der Waals forces on a finite number $N$ of suspended particles. The macroscopic limit does only exist for the spectral force (Fig.~\ref{fig:spectral}) and the spectral stresses $\widetilde{\sigma}_F$, but not for the total force density $f$ of Eq.~(\ref{eq:fresult}) nor for the total stresses integrated over $\kappa$. Figure~\ref{fig:asymptotics} compares the spectral stresses in the macroscopic limit with the asymptotics (\ref{eq:asysigma}) for large wavenumbers, confirming our result (\ref{eq:asysigmatotal}). Figure~\ref{fig:renorm} illustrates the main result of this paper, the renormalized electromagnetic stresses, evaluated according to Eq.~(\ref{eq:eff}) for the example considered here. 

\section{Conclusions}

We have analyzed a simple model for the Casimir force in dielectrics: a chain of point particles interacting with each other by scattering virtual electromagnetic waves. The scattering creates forces \cite{Buhmann,Rodriguez,Forces} on the particles that appear as the familiar van der Waals forces \cite{ShahmoonvdW} (with retardation taken into account \cite{CasimirPolder}). The point scatterers may represent the molecules of a dielectric medium \cite{LL8} and they do represent colloidal suspensions \cite{Israelachvili} such as liquid paint. In our model, we assume both the particles and the field to be confined to one dimension, which has allowed us to develop an efficient numerical algorithm for calculating the forces, and to deduce the macroscopic limit in unprecedented detail and clarity. 

In the macroscopic limit of infinitely many, infinitely weak scatterers, we found a result that seems both satisfying and paradoxical: the van der Waals force becomes the gradient force of electric field fluctuations, which nicely connects the scattering approach to Casimir forces \cite{Lambrecht,Wirzba,Ingold} with macroscopic quantum electrodynamics \cite{Buhmann,KSW} and the rich field of optical forces \cite{Ashkin1,Ashkin2,Ashkin3,Neuman,Pesce}. If the scatterers are arranged to form a piece--wise homogeneous medium the force density is finite, except at interfaces. Yet any inhomogeneity, however minute, will produce infinite forces and bring about the immediate collapse of the suspension. The scatterers appear to anticipate whether they are aligned to form pieces of a homogeneous or of an inhomogeneous medium, which seems paradoxical. It also disagrees with the empirical fact that some colloidal suspensions are stable --- liquid paint may remain liquid even if it is not perfectly smooth. Homogeneous colloidal suspensions are known to be stable when the long--range attractive van der Waals forces are compensated by other short--range repulsive forces \cite{Israelachvili}.

We have argued that the same mechanism that makes homogeneous colloidal suspensions stable does also prevent the collapse of inhomogeneous media. The counter forces compensate the local contributions to the van der Waals stress, and the remaining non--local contributions constitute the Casimir force. This local balance between the van der Waals and the counter force amounts to the renormalization of the quantum--electromagnetic stresses. We have shown [Eqs.~(\ref{eq:sigma0}) and (\ref{eq:eff})] that the counter stress depends on the local macroscopic dielectric response functions, the refractive index and its first and second derivative, independent of the microscopic physics involved. 

Our model describes a suspension of particles in a fluid, but we could also apply it to solids where the inter--atomic forces are even stronger. In either case, the result of the counter pressure is independent of the material details.  Only the refractive--index profile counts. Suppose we replace the dielectric material by empty space. Even in this case we need to remove the local stress. For example (Sec.VIA) in each homogenous layer we should remove $n\kappa$ in the spectral stress [Eq.~(\ref{eq:stresses})] --- also in empty space where $n=1$. Only then the discontinuity of the stress gives the correct Casimir--Lifshitz force \cite{Lifshitz} [Eq.~(\ref{eq:lifshitz})]. One might speculate whether there is some mechanism of counter forces in empty space as well, acting not on the molecular scale of $10^{-9}\mathrm{m}$, but rather on the Planck scale of $10^{-35}\mathrm{m}$. The precise microscopic nature of such a space--time medium is unknown, but as in ordinary dielectric materials, the renormalization should not depend on microscopic details. 

In curved space--time, {\it i.e.}\ in gravitational fields \cite{LL2} the relation between space and media is well--known \cite{Gordon,Plebanski,Schleich,LeoPhil} in classical electromagnetism: there, Maxwell's equations are the equations of electromagnetic fields in impedance--matched magneto--electric media \cite{Plebanski}. In our model of quantum electromagnetism in media, we have reproduced an intriguing feature of renormalization in curved space--time, the van der Waals anomaly (Sec.~VC). This dielectric analogue of the trace anomaly \cite{Wald} of quantum fields in curved space \cite{BD} has been derived before in spherical and planar media \cite{Itai}, here we have deduced it in the most elementary way. In general relativity, the trace anomaly appears in Einstein's field equations \cite{LL2} as the cosmological constant \cite{LondonProceedings}, it has the correct order of magnitude and is consistent with astronomical data \cite{Berechya}, so perhaps it simply {\it is} the cosmological constant \cite{Annals,Tkatchenko}. 

Our simple model has come a long way. It has shown to contain rich physics ranging from liquid paint to the cosmological constant. The model is simple enough for elementary calculations, but sufficiently sophisticated for deep cross--field connections. We hope to have made it ``as simple as possible, but not simpler''.

\section*{Acknowledgements}

We thank
Yael Avni,
Gregory Falkovich,
Jonathan Kogman,
Ziv Landau,
Alexander Poddubny,
and
Ren\'{e} Sedmik
for discussions.
L.M.R. was supported by the Austrian Science Fund (FWF) through Project No. P 32300-N27 (WaveLand).
U.L. was supported by the Israel Science Foundation and the Murray B. Koffler Professorial Chair. 
For calculating the quantum forces between large numbers of particles we used the Vienna Scientific Cluster (VSC).

\appendix

\section{Numerical procedures}

Our method of using the transfer matrix for calculating Casimir forces on individual scatterers is efficient and fast, but two problems arise for large wavenumbers $\kappa$  (becoming relevant when the number $N$ of scatterers is large): the products of the exponential propagation matrices (\ref{eq:pmatrix}) in the recurrence relation (\ref{eq:recurrence}) may cause overflow and the required $\kappa$ integrations in the force formula (\ref{eq:forceformula}) may be slow or insufficiently accurate. We recommend the remedies described here. We also indicate how the calculations can be done in multiple parallel threads. 

But first we make the following simplification. We do not need the full transfer matrix and its derivative, but only the elements $T_{22}$ and $\partial_jT_{22}$ for calculating the force [Eq.~(\ref{eq:forceformula})]. Equations~(\ref{eq:TAB}) and (\ref{eq:dT22}) and the recurrence relations (\ref{eq:recurrence}) imply that both $T_{22}$ and $\partial_jT_{22}$ depend only on the second column of A and the second row of B. We may thus replace the A and B matrices by $N$ two--dimensional vectors:
\begin{equation}
{\bm a}_j = 
\begin{pmatrix}
A_{12} \\
A_{22} 
\end{pmatrix}
\,,\quad
{\bm b}_j = 
\begin{pmatrix}
B_{21} \\
B_{22} 
\end{pmatrix}
.
\end{equation}
We may give the ${\bm a}$ and ${\bm b}$ the initial values
\begin{equation}
{\bm a}_1 = {\bm b}_N = 
\begin{pmatrix} 0\\ 1 \end{pmatrix}
\end{equation}
as the $e^{\kappa\delta}$ factors in the initial conditions (\ref{eq:initial}) drop out in the ratio $T'_{22}/T_{22}$. Physically this means that the force is independent of the propagation to and from the chain of scatterers.

\subsection{Recurrence}

From the macroscopic limit (Sec.~IIIA) we know that the ${\bm a}$ and ${\bm b}$ vectors approach the structure (\ref{eq:structure}) and go with solutions of the Helmholtz equation (\ref{eq:helmholtz}). Furthermore, from the Madelung representation (\ref{eq:madelung}) with asymptotics (\ref{eq:kasy}) follows that ${\bm a}$ and ${\bm b}$ go as the exponential of $\int \!n\kappa\,dx$, which may grow exponentially large, causing overflow. However, as the force [Eqs.~(\ref{eq:forceformula}) and (\ref{eq:det})] depends only on the ratio between $T'_{22}$ and $T_{22}$ we may multiply the ${\bm a}$ and ${\bm b}$ vectors by some factor, and get the same result. To avoid overflow, we recommend to employ the inverse of the macroscopic propagation factor as compensation. This means one should replace the recurrence relations (\ref{eq:recurrence}) by
\begin{gather}
{\bm a}_{j+1}= e^{-n\kappa \delta}\, {\rm P}\, {\rm R}_j\, {\bm a}_j \,,\quad
{\bm b}_{j-1} = e^{-n\kappa \delta}\, {\rm P}\, {\rm R}_j^\top {\bm b}_j 
\nonumber\\
\mbox{with}\quad n= \sqrt{1+\chi_j} 
\label{eq:recurrence1}
\end{gather}
where the ${\rm P}$ and ${\rm R}_j$ denote the matrices
\begin{equation}
{\rm P} = \begin{pmatrix}
e^{-\kappa\delta} & 0 \\
0 & e^{\kappa\delta} 
\end{pmatrix}
,\quad
{\rm R}_j = \mathbb{1} - \frac{\kappa\delta \chi_j}{2}
\begin{pmatrix}
1 & 1 \\
-1 & -1
\end{pmatrix}
.
\end{equation}
We obtain from Eqs.~(\ref{eq:TAB}) and (\ref{eq:dT22}) for the $j$--th particle:
\begin{equation}
T_{22} = {\bm b}_j \cdot {\rm R}_j \, {\bm a}_j
\,,\quad
\partial_jT_{22}= -\delta \kappa^2 \chi_j\,  {\bm b}_j\cdot \sigma_x {\bm a}_j
\label{eq:T22s}
\end{equation}
in terms of the $\sigma_x$ Pauli matrix defined in Eq.~(\ref{eq:pauli}). Note that for the calculation of the force on one particle the required transfer--matrix elements for all other particles are computed. It is wise to cache them such that they may be used in the force integral for other particles as well. 

\subsection{Integration}

For determining the force on the $j$--th particle we need to compute the integral
\begin{equation}
F_j=-\frac{\hbar c}{2\pi} \int_0^\infty  \frac{\partial_jT_{22}}{T_{22}} \,d\kappa \,.
\label{eq:forceformula1}
\end{equation}
For numerical integration, we use a modified variant of the Romberg integration method \cite{NumRec} of order $2 \times 5$. The Romberg method converges much faster than many other numerical integration algorithms, and it also has the advantage that more and more function evaluations in--between the already evaluated values can be added in each iterative step, until the required precision has been reached. For integrating from zero to infinity, we apply the  ``Mixed Rule'' transformation \cite{NumRec} $\kappa = \exp[\nu-\exp(-\nu)]$ turning Eq.~(\ref{eq:forceformula1}) into an integral from $-\infty$ to $+\infty$ that needs to be cut off.  Both cutoff values of $\nu$ are determined by an adaptive search algorithm. Further, to make use of the caching--method mentioned above (Sec.~A1) we have modified the algorithm such that all possible floating--point values of $\nu$ required for the Romberg integration get unique integer values assigned. These values serve as lookup index for the cache. For checking the accuracy of our calculations we used arbitrary--precision arithmetics packages. 

\subsection{Parallelization}

In our software implementation, multiple of the above--described routines are executed in parallel threads. This has to be considered in the caching mechanism, which not only must be thread--safe, but also must be able to deal with the possibility that a requested result is in the limbo state of being currently in calculation by another thread. Multi--threading has allowed us to run accurate calculations of the van der Waals force on millions of particles.

\section{Previous renormalization}

How is the renormalization procedure of this paper (Sec.~V) related to the previous one \cite{Itai}? In the previous procedure \cite{Itai}, an outgoing Green function is subtracted from the full Green function as well, but this Green function is not given as the exact solution of the Green equation (\ref{eq:greeneq}) with appropriate conditions at the source, Eqs.~(\ref{eq:out}) and (\ref{eq:dpm}). Rather it comes from the geometrical--optics approximation of the original Green function [Eq.~(\ref{eq:green})] in the vicinity of the source where, in the renormalizer, $n(x)$ is taken as the quadratic expansion
\begin{equation}
n(x) = n_0 + n_0'(x-x_0) + \frac{n_0''}{2}(x-x_0)^2 
\label{eq:quadratic}
\end{equation}
with $n_0=n(x_0)$. We have confirmed [Eqs.~(\ref{eq:beta0})] and (\ref{eq:asysigma})] that the local stresses do only depend on maximally the second derivative of $n$, which justifies the quadratic expansion (\ref{eq:quadratic}). In our case, this is not an assumption, but a consequence of the boundary conditions (\ref{eq:out}) and (\ref{eq:dpm}). 

The outgoing Green function $g_0$ is represented in terms of the optical length $s$, the amplitude ${\cal A}$ and the first-scattering coefficient $\beta_1$ as
\begin{equation}
g_0 = \mathrm{e}^{-\kappa s} {\cal A} \left(1+\frac{\beta_1 s}{\kappa}\right) .
\end{equation}
The optical length is a solution of the eikonal equation \cite{Renormalization} (in one dimension $s'^2=n^2$) as $(x-x_0)$ times a term up to quadratic order. Integrating the eikonal equation we obtain
\begin{equation}
s=(x-x_0)\left[n_0 + \frac{n_0'}{2}(x-x_0) + \frac{n_0''}{6}(x-x_0)^2\right]
.
\label{eq:s}
\end{equation}
The amplitude satisfies the conservation law \cite{Renormalization} $({\cal A}^2 s')'=0$ (in one dimension). We get by integration:
\begin{equation}
{\cal A} = -\frac{1}{2\kappa \sqrt{n n_0}}
\end{equation}
with $n$ given by the quadratic expansion (\ref{eq:quadratic}). The integration constant gives the prefactor of the amplitude. It is chosen such that $g_0$ satisfies the Green equation (\ref{eq:green}) in the immediate vicinity of the delta function where $n$ can be taken as constant $n_0$. The first--scattering coefficient \cite{Renormalization} $\beta_1$ obeys $2n \beta_1 ={\cal A}''/{\cal A}$ evaluated at $x_0$. We express it in terms of the $\beta_0$ defined by Eq.~(\ref{eq:beta0}) and obtain:
\begin{equation}
\beta_1 = \frac{\beta_0}{n_0} \,.
\end{equation}

We substitute these expressions in Eq.~(\ref{eq:sigmas0}) for the stresses. Up to $\kappa^{-1}$ order we obtain exactly the same result [Eq.~(\ref{eq:asysigma})] as with the renormalizer advocated in this paper, including the need for the anomaly. The differences lie in the convergent part, but it is the convergent part of $\sigma-\sigma_0$ that gives the actual physical force. Another difference is conceptual: the new procedure is much more clear and elegant. However, it would be very useful to test the theory in experiments, including computer experiments, bearing in mind the theorem: for every difficult problem there is a solution that is simple, elegant, and wrong. 

\section{Three layers}

In this appendix we consider the case of three homogeneous dielectric layers (as in Sec.~VIA) and derive formulas (\ref{eq:gmiddle}-\ref{eq:gleft}) for the Green function. Expressions for this Green functions are scattered in the literature, with inconsistent notations, and not all what we need can be found at one place. For the convenience of the reader we derive them here with an efficient method, the transfer--matrix method (Sec.~IIB).

Consider the transfer from the one homogenous layer with refractive index $n_1$ to another layer with index $n_2$. In order to satisfy the Helmholtz equation (\ref{eq:helmholtz}) the field $\psi$ and its first derivative $\psi'$ must be continuous. From this follows the transfer matrix ${\rm T}_l$ across the interface:
\begin{equation}
{\rm T}_l =
\frac{1}{2n_2}
\begin{pmatrix}
n_2+n_1& n_2-n_1 \\
n_2-n_1 & n_2+n_1
\end{pmatrix}
.
\label{eq:TL}
\end{equation}
In our case, it describes the transfer across the left interface of the central layer. We will also need the transfer matrix at the right interface:
\begin{equation}
{\rm T}_r =
\frac{1}{2n_3}
\begin{pmatrix}
n_3+n_2& n_3-n_2 \\
n_3-n_2 & n_3+n_2
\end{pmatrix}
.
\label{eq:TR}
\end{equation}
We express the Green function as the product (\ref{eq:green}) of wave functions $\psi_\pm$ divided by their Wronskian (\ref{eq:wronskian}). The $\psi_\pm$ we determine with transfer matrices.

Consider first the case where both the point of emission $x_0$ and the observation point $x$ lie in the central layer. There we make the ansatz $\psi_+=e^{n_2\kappa x}+\varrho \,e^{-n_2\kappa x}$. In the left layer $\psi_+\propto e^{n_1\kappa x}$. Consequently,
\begin{equation}
{\rm T}_l 
\begin{pmatrix}
0 \\
1
\end{pmatrix}
\propto
\begin{pmatrix}
\varrho \\
1
\end{pmatrix}
,
\end{equation}
which gives for $\varrho$ the Fresnel coefficient $\varrho_l$ of Eq.~(\ref{eq:fresnel}). For $\psi_-$ we make the ansatz $\psi_-=e^{-n_2\kappa x}+\varrho \,e^{n_2\kappa x}$ in the central layer and require $\psi_-\propto e^{-n_3\kappa x}$ after propagating to the right layer and crossing the interface:
\begin{equation}
{\rm T}_r 
\begin{pmatrix}
e^{-n_2\kappa a} & 0\\
0 & e^{n_2\kappa a} 
\end{pmatrix}
\begin{pmatrix}
1 \\
\varrho
\end{pmatrix}
\propto
\begin{pmatrix}
1 \\
0
\end{pmatrix}
,
\end{equation}
which gives $\varrho = \varrho_r\,e^{-2n_2\kappa a}$ in terms of the other Fresnel coefficient (\ref{eq:fresnel}). We thus have in the central layer:
\begin{eqnarray}
\psi_+ &=& e^{n_2\kappa x}+\varrho_l \,e^{-n_2\kappa x} 
\,,\nonumber\\
\psi_-&=&e^{-n_2\kappa x}+\varrho_r \,e^{n_2\kappa (x-2a)} \,.
\label{eq:psimiddle}
\end{eqnarray}
The Wronskian (\ref{eq:wronskian}) we read off from Eq.~(\ref{eq:psimiddle}) as
\begin{equation}
W = -2n_2\kappa \left(1-\varrho_l\varrho_r\,e^{-2n_2\kappa a}\right) .
\label{eq:Wmiddle}
\end{equation}
Expressions (\ref{eq:psimiddle}) and (\ref{eq:Wmiddle}) give the Green function (\ref{eq:green}). Note that one can interpret this $g(x,x_0)$ as the summed--up geometrical series of all multiple reflections \cite{Essential}. In Sec.~VIA we need $g(x,x)=W^{-1}\psi_+(x)\psi_-(x)$ and get Eq.~(\ref{eq:gmiddle}) from the expressions derived here.

Now turn to the left layer. There $\psi_+$ naturally propagates without scattering as $e^{n_1\kappa x}$ while $\psi_-$ is a superposition of the $e^{\pm n_1\kappa x}$. Therefore we write
\begin{equation}
\psi_+ = e^{n_1\kappa x} \,,\quad
\psi_-=e^{-n_1\kappa x}+\varrho\,e^{n_1\kappa x} 
\label{eq:psileft}
\end{equation}
and have the Wronskian
\begin{equation}
W = -2n_1\kappa \,.
\label{eq:Wleft}
\end{equation}
We require that $\psi_-\propto e^{-n_3\kappa x}$ after arriving at the right layer:
\begin{equation}
{\rm T}_r 
\begin{pmatrix}
e^{-n_2\kappa a} & 0\\
0 & e^{n_2\kappa a} 
\end{pmatrix}
{\rm T}_l
\begin{pmatrix}
1 \\
\varrho
\end{pmatrix}
\propto
\begin{pmatrix}
1 \\
0
\end{pmatrix}
,
\label{eq:trans}
\end{equation}
solve for $\varrho$ and express the result in terms of the Fresnel coefficients (\ref{eq:fresnel}):
\begin{equation}
\varrho = - \frac{\varrho_l-\varrho_r e^{-2n_2\kappa a}}{1-\varrho_l\varrho_r\,e^{-2n_2\kappa a}} 
\,.
\label{eq:rho}
\end{equation}
Inserting the result (\ref{eq:rho}) in expressions (\ref{eq:psileft}) for the waves $\psi_\pm$ with Wronskian (\ref{eq:Wleft}) we obtain the desired Green function $W^{-1}\psi_+(x)\psi_-(x)$ and arrive at formula (\ref{eq:gleft}).

The coefficient (\ref{eq:rho}) describes the reflection at the entire central layer. The physical interpretation of $\varrho$ becomes apparent when we mathematically transform formula (\ref{eq:rho}) to
\begin{equation}
\varrho = -\varrho_l + (1-\varrho_l^2)\,\varrho_r\,\frac{e^{-2n_2\kappa a}}{1-\varrho_l\varrho_r\,e^{-2n_2\kappa a}} 
\label{eq:rho1}
\end{equation}
and employ the geometric series
\begin{equation}
\frac{1}{1-\varrho_l\varrho_r\,e^{-2n_2\kappa a}} = \sum_{m=0}^\infty \left(\varrho_l\varrho_r\,e^{-2n_2\kappa a}\right)^m \,.
\label{eq:series}
\end{equation}
Equations (\ref{eq:rho1}) and (\ref{eq:series}) describes the following scenario: a wave comes in from the left. At the left interface a part of the wave is reflected with reflectivity $-\varrho_l$ (the minus sign because $\varrho_l$ describes reflection from the right to the left, not from the left to the right). The other part is transmitted with transmittance $\sqrt{1-\varrho_l^2}$. The transmitted part propagates to the right interface with factor $e^{-n_2\kappa a}$. If the wave makes it back (such that it counts in the overall reflection coefficient $\varrho$) it must get reflected at the right interface with coefficient $\varrho_r$ and propagate back to the left interface with factor $e^{-n_2\kappa a}$ and be transmitted out of the central layer with transmittance $\sqrt{1-\varrho_l^2}$. This gives the factor $(1-\varrho_l^2)\,\varrho_r\,e^{-2n_2\kappa a}$ in Eq.~(\ref{eq:rho1}). The wave may also get multiply reflected in the central layer, bouncing back and forth multiple times, as the series (\ref{eq:series}) describes. 
 
,


\begin{thebibliography}{99}
\bibitem{Remark}
Note that this simple substitution relies on the fact that $K+i\Gamma$ is analytic and decaying on the lower half plane of complex time difference, which is only true for the vacuum state. In the thermal state $K$ is periodic (Kubo--Martin--Schwinger relation \cite{Annals,LL7}).
\end{thebibliography}
\end{document}